# Coupled fiber taper extraction of 1.53 um photoluminescence from erbium doped silicon nitride photonic crystal cavities

**Gary Shambat[1,*], Yiyang Gong[1], Jesse Lu[1], Selçuk Yerci[2], Rui Li[2], Luca Dal Negro[2], and Jelena Vučković[1]**

[1]*Department of Electrical Engineering, Stanford University, Stanford 94305*
[2]*Department of Electrical and Computer Engineering, Boston University, Boston, MA 02215*
[*]*email:gshambat@stanford.edu*

**Abstract:** Optical fiber tapers are used to collect photoluminescence emission at ~1.5 um from photonic crystal cavities fabricated in erbium doped silicon nitride on silicon. Photoluminescence collection via fiber taper is enhanced 2.5 times relative to free space, with a total taper collection efficiency of 53%. By varying the fiber taper offset from the cavity, a broad tuning range of coupling strength is obtained. This material system combined with fiber taper collection is promising for building on-chip optical amplifiers.

___________________________________________________________________________

______________________________________________________________________________

## 1. Introduction

Optical interconnects have the potential to improve data communication in computer systems by providing a high bandwidth, low power alternative to present electrical interconnects. Particularly for off-chip links, where traditional metal interconnects suffer from large capacitive delays and high energy consumption, implementing optical interconnects may be very advantageous [1]. While much progress has been made with CMOS-compatible modulators and detectors [2-5], a suitable light source that is well integrated with a silicon platform remains to be demonstrated. Recently, it was shown that photonic crystal (PC) cavities in erbium-doped silicon nitride (Er:SiN$_x$) exhibited linewidth narrowing with increased optical pump power [6, 7], demonstrating that 31% of Er ions can be excited under optical pumping. Such a material may be suitable for an on-chip amplifier or laser source in the future.

To connect a nanoscale photonic crystal cavity to the off-chip environment of an optical interconnect system, silica fiber tapers may be used to in- and out-couple light efficiently. Fiber tapers have been shown to be useful not only in probing and characterizing cavities [8,9], but also in providing optical pumping for a laser and extracting photoluminescence [10]. Because fiber tapers rely on resonant evanescent coupling, they can inject and extract light with much higher efficiency than with free space techniques. In this study, we report the extraction of photoluminescence from Er:SiN$_x$ PC cavities into fiber tapers with much higher efficiency than outcoupling through free space. We also demonstrate that a broad tuning range of quality factor (up to 98% of the intrinsic Q value) is attainable by altering the strength of the coupling and thus the coupling efficiency.

## 2. Fabrication

*2.1 Taper fabrication:*

Fiber tapers were fabricated using the well known "flame brushing" technique [11] in which a standard single mode communication fiber is simultaneously heated by a torch and pulled outward by motorized stages. A hydrogen-oxygen torch was used to provide a small flame that was quickly scanned back and forth along the designated "hot zone" of the fiber. In order to provide the best mechanical stability of the taper probe, the pull length was kept to only ~3 mm, while the taper diameter was ~1 um. This diameter was chosen because when the taper is below approximately 1.1 um the waveguide becomes single mode [12]. This single mode behavior cleans up coupling spectra and removes erroneous fringing in the transmission and photoluminescence spectra. After the fiber tapers were drawn, they were carefully bent such that the radius of curvature was small enough for the tapers to interact only with photonic crystal cavities and not the nearby substrate. An optical image of a fabricated taper is shown in Fig. 1(a). The whole fiber was mounted to a glass slide and was fusion spliced into connector cables. Finally, the slide was mounted in a measurement rig where the taper could be positioned onto photonic crystal cavities via both motorized and manual nanopositioning stages. The transmission of the mounted fiber taper was 16%, with most of the loss coming from scattering within the short taper transition regions.

*2.2 Cavity fabrication:*

Photonic crystal structures were fabricated in a two layer membrane consisting of a bottom silicon layer and a top, light emitting, erbium-doped silicon nitride (Er:SiNx) layer, same as in our previous work [7]. Approximately 110 nm of erbium-doped amorphous silicon nitride was deposited on a SOI wafer with a silicon thickness of 250 nm by $N_2$ reactive magnetron co-sputtering from Si and Er targets in a Denton Discovery 18 confocal-target sputtering system [13].

Photonic crystal L3 cavities [14] were patterned with an electron beam lithography system using ZEP-520A as the resist. A $Cl_2$:HBr dry etch chemistry was used to etch the air holes and the sacrificial oxide layer in the SOI was undercut at the end using a 6:1 Buffered oxide etch to form the suspended PC membrane. The lattice periodicity was set to a = 410 nm, and hole radius was r = 125 nm. Additional outward shifts of 0.15a and 0.075a for the on-axis holes surrounding the cavity were made to increase the intrinsic Q of the cavity. Fig. 1(b) shows a scanning electron microscope (SEM) image of the tested cavity.

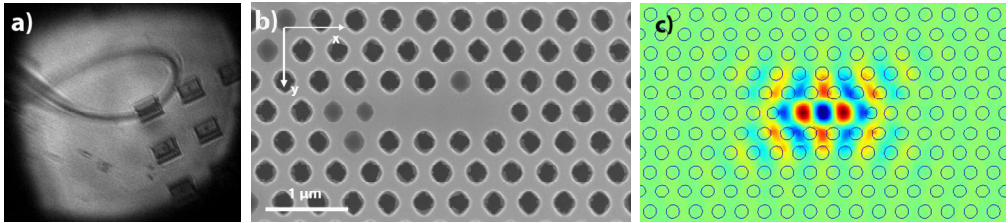

Fig. 1. (a) Optical microscope image of the fiber taper. (b) SEM picture of the fabricated $SiN_x$/Si photonic crystal cavity. (c) Calculated $E_y$ field of the fundamental mode of the photonic crystal.

## 3. Theoretical analysis

*3.1 – Free space collection efficiency*

Finite difference time domain (FDTD) calculations were performed on the cavity structure with the size parameters of the SEM image above. The computational grid contained 10 mirror periods on each side of the cavity in the y direction and 12 mirror periods on each side in the x direction. Q-factor calculations were performed by integrating the power flux through the appropriate boundaries: top and bottom half spaces for the vertical Q's and slab walls for the parallel Q. The fundamental mode profile is shown in Fig. 1(c).

In order to estimate the percentage of light collected by our microscope objective, a two-dimensional Fourier Transform was performed on the field components of the cavity mode just above the slab. From the k-space distribution of the tangential electric and magnetic fields, the total power radiated within a 0.5 numerical aperture (NA) objective was determined [15] to be 10% of the total emission into the half space above the slab. The low value is not surprising if one examines the k-space map for the cavity $E_y$ field shown in Figure 2. The transverse electric-like (TE-like) fundamental mode is predominantly $E_y$ in character and from the k-space pattern in Fig. 2(b), we see that the magnitude of the components radiated within the NA cone is much smaller than the sum inside the light cone. Similar observations were made for the other tangential fields. Additionally, since the bottom half of the slab radiates approximately 1.4 times more than the top half, as indicated by the top and bottom Q factors of the FDTD simulation (this is a result of the PC membrane asymmetry), the total efficiency of collection of light into the objective is approximately 4.2%.

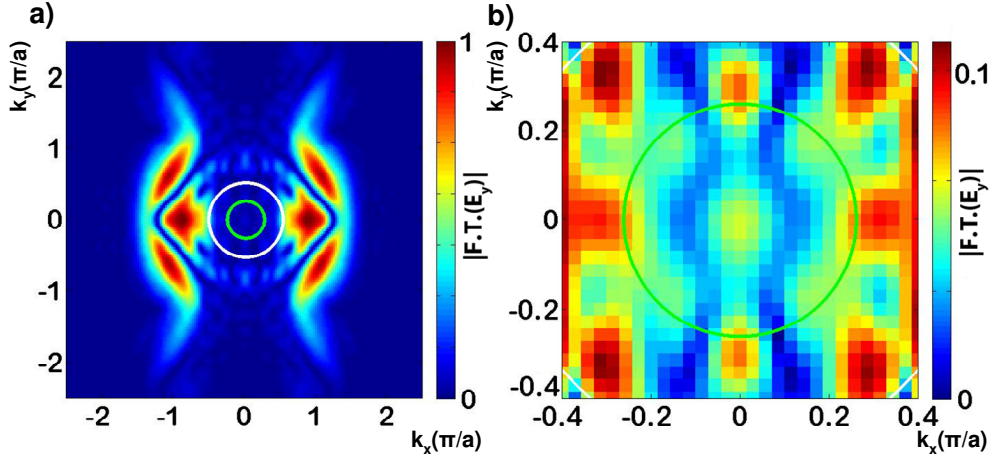

Fig. 2. (a) K-space map for the $E_y$ component of the fundamental mode, taken just above the PC slab. The white circle is the light cone while the green circle is the NA = 0.5 cone. (b) Zoomed in image of the previous plot highlighting the difference between the radiated components captured and not captured by the objective lens. The magnitude of the k components inside the NA = 0.5 cone is clearly much lower than those outside the cone.

*3.2 – Coupled cavity Q-factor*

The intrinsic cavity Q for the fundamental mode was found to be 16,000 and is dominated by the in-plane Q which was only 26,000. Due to asymmetry of the PC in the vertical direction, the confined TE-like cavity mode couples to TM-like propagating Bloch states, which suffer from an incomplete bandgap and hence leak out of the cavity [16]. In contrast, a similar PC structure composed of just the symmetric silicon layer has an in-plane Q of nearly 900,000 and a net Q of 61,000.

Fig. 3 shows a plot of the various FDTD simulated Q-factors as a fiber taper, represented as a cylinder of silica (n = 1.45, diameter = 1 um), is scanned across the surface of the L3 cavity along the y-axis. For zero offset, we see an expected drop in the top Q due to the additional decay into the fiber. The in-plane Q is also reduced by a factor of two due to the increased vertical asymmetry of the structure. This parasitic decay reduces the overall Q beyond the intrinsic limit of just additional decay into the fiber.

As the taper is shifted to the side of the cavity, several trends can be observed. First, the top Q is seen to reach a minimum for an offset of 250 nm, before increasing monotonically towards the intrinsic limit of 104,000. The decrease in coupling strength due to a smaller field overlap for the offset taper causes the vertical decay to decrease as expected. The bottom Q is mostly unaffected by the taper but does show some increased decay when the taper is in close proximity to the cavity. The in-plane Q returns to its original value as the taper is moved farther away from the cavity as well. Finally, we see that even for large offsets, the fiber taper can still couple with the cavity without inducing parasitic decay. In particular, at 1.3 um offset, the in-plane and bottom Q are at their intrinsic limits while the top Q is 75,000, below the intrinsic value of 104,000. As a rough estimate of the collection efficiency when the taper offset is zero, if the top Q is composed of only the intrinsic vertical decay and the new fiber-induced decay, then the fiber Q is ~16,000. Meanwhile the total Q for zero offset is ~7,000 and thus the collection efficiency, or ratio of decay into fiber versus total decay, is 7,000/16,000 ≈ 44%. Therefore simulations predict that a much larger (~10-fold) coupling efficiency can be achieved into the fiber relative to the coupling to free space and a collection lens of NA = 0.5.

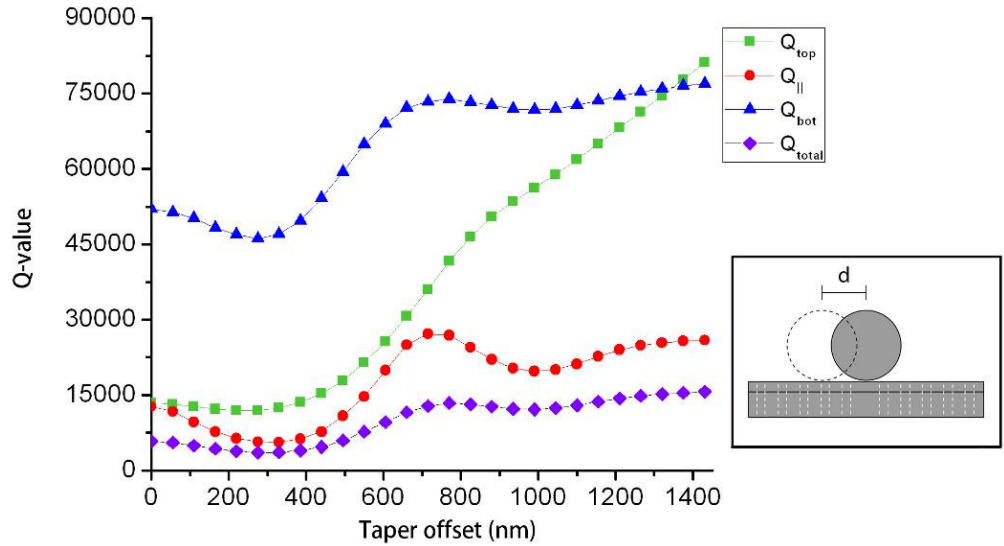

Fig. 3. Variation of the top, in-plane, bottom, and total Q factors as a function of taper displacement, d, along the y axis. When the taper is directly over the cavity, the vertical Q goes down expectedly and the in-plane Q is parasitically reduced by a factor of two. For large taper offsets, the Q-values approach their intrinsic limits. Inset shows the cross-section of the FDTD simulated structure.

## 4. Experiment

*4.1 – Free space photoluminescence*

Photoluminescence (PL) from fabricated photonic crystal cavities was collected in free space, in the direction perpendicular to the PC plane, through a 100x (NA = 0.5) objective lens, directed toward a monochromator, and detected with a liquid nitrogen cooled linear InGaAs CCD array, same as in our earlier work [7]. PL from the cavity region was spatially filtered with an iris in order to minimize the background signal and optimized for greatest collection efficiency. Samples were pumped with a 980 nm laser diode at 6 mW. Fig. 4 displays the PL of the L3 cavity showing the fundamental resonance at approximately 1535.2 nm at the maximum of erbium's typical emission spectrum. This particular cavity was chosen for testing because its cavity resonance overlaps with the peak emission of erbium. A higher order mode is observed at 1480 nm and the features at 1560 nm are likely band edge modes [17]. The measured Q value of the intrinsic, unloaded cavity was found to be 12,630 as seen by the fit to a Lorentzian in the inset of Fig. 4.

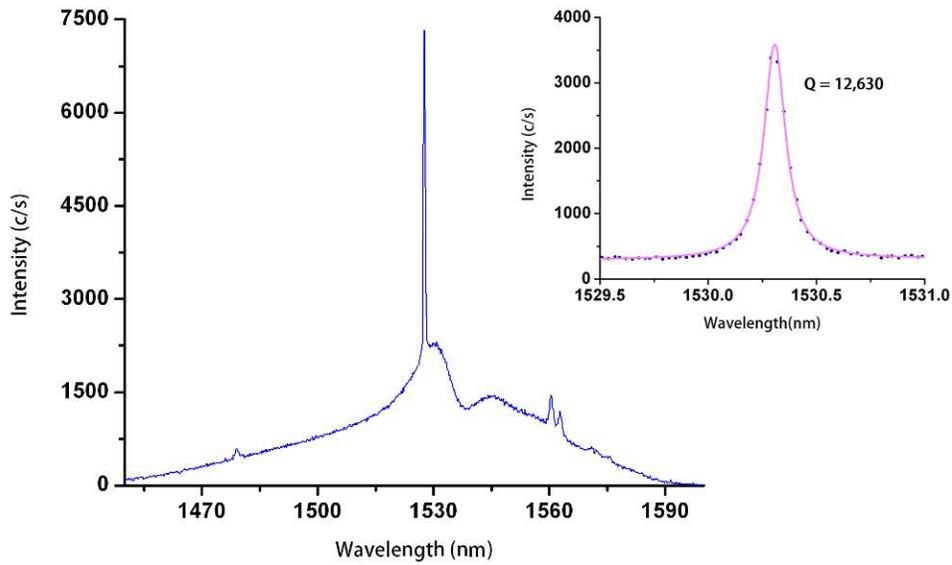

Fig. 4. PL spectrum of the Er:SiNx on Si PC cavity, collected from free space in the direction perpendicular to the PC plane (through an objective lens with NA = 0.5). The fundamental cavity mode corresponds to the highest peak near 1530nm, and the background emission has the expected erbium profile. The inset shows the fundamental peak zoomed in with a Lorentzian fit.

*4.2 – Passive (transmission) cavity measurements via fiber taper*

Transmission measurements were performed by coupling a broadband source into a fiber taper that was well aligned with an L3 cavity main axis, and by measuring the spectrum at the output of the fiber with an optical spectrum analyzer (OSA) (Fig. 5(a)). Light was not polarized at the input so that all the modes of the cavity could be excited, thus facilitating the identification of the fundamental mode. The taper rested on the sample surface.

Fig. 5(b) shows a full transmission spectrum of the cavity whose PL is shown in Fig. 4. The fundamental resonance is clearly seen at about 1530 nm, as in Fig. 4, as well as two higher order modes at shorter wavelengths. Four additional modes were found at wavelengths below 1400 nm. The background loss of 2-5 dB comes primarily from TM loss in the PC slab and characteristic absorption of the erbium-doped nitride material.

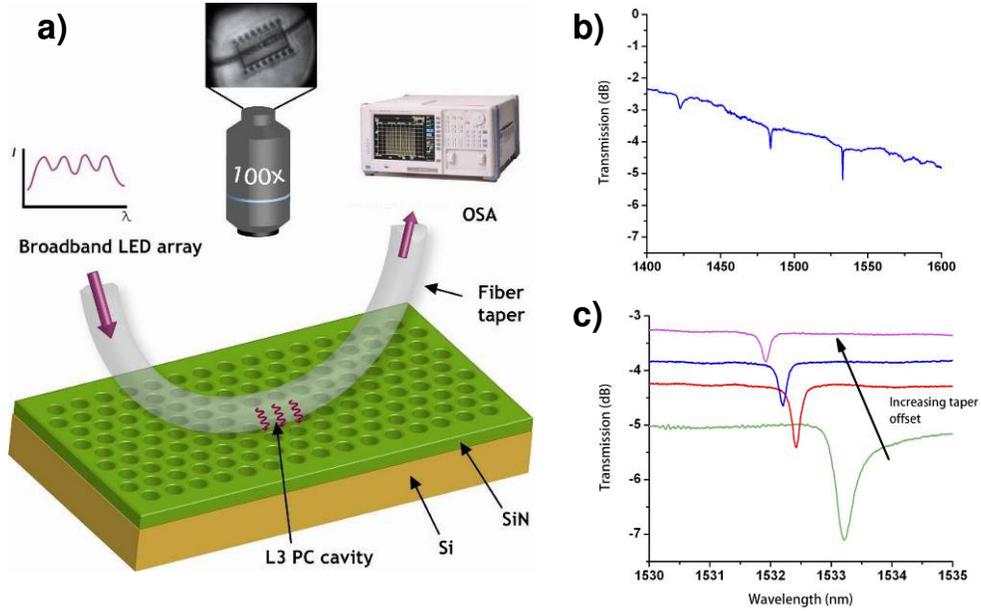

Fig. 5. (a) Experimental setup for measurement of transmission through the Er:SiNx on Si PC cavities via fiber taper. The fiber taper is aligned and in contact with the L3 cavity. The output spectrum of a broadband source at the input of the fiber is measured by an OSA. Inset shows optical picture of aligned taper. (b) Full transmission spectrum of the L3 cavity. (c) Multiple transmission spectra for the same cavity but with different fiber taper offsets along the y axis.

In Fig. 5(c), multiple transmission scans are taken for the taper as it is progressively offset from the main axis in of the cavity in the y-direction. The short pull length of the taper was necessary for the tension of the fiber to be large enough to allow this seamless translation, which is otherwise prevented by the Van der Waals sticking attraction between the fiber and semiconductor. As can be seen in the plot, as the fiber taper is dragged away from the cavity center, the transmission coupling magnitude drops and the Q factor increases. This is not surprising since the coupling coefficient is given by the overlap between the cavity and waveguide modes [18]. As the taper shifts, this overlap decreases along with the transmission coupling. Likewise, since this secondary decay path via the taper decreases with taper offset, the measured Q value approaches the intrinsic limit. Also worth noting is the red shift induced by the taper as it moves over the cavity which is due to the increased effective index of the cavity.

### 4.3 – Fiber-outcoupled photoluminescence

Fiber taper coupling provides a convenient means to extract and collect photoluminescence from a cavity mode that would ordinarily radiate non-uniformly into free space. A well designed cavity (such as the described L3 type) is meant to have few wave vector components in the light cone in order to maximize the Q (see Fig. 2) [14]. The fiber taper can, on the other hand, extract those components that are outside the light cone through evanescent coupling.

PL measurements with a coupled fiber taper were taken under similar conditions as those for the free space measurement. To do so, a 980 nm laser diode pump with power 6 mW was focused to a small spot where the fiber taper rested on the cavity. The output from one end of the fiber was then sent to the same spectrometer setup as before (Fig. 6(a)).

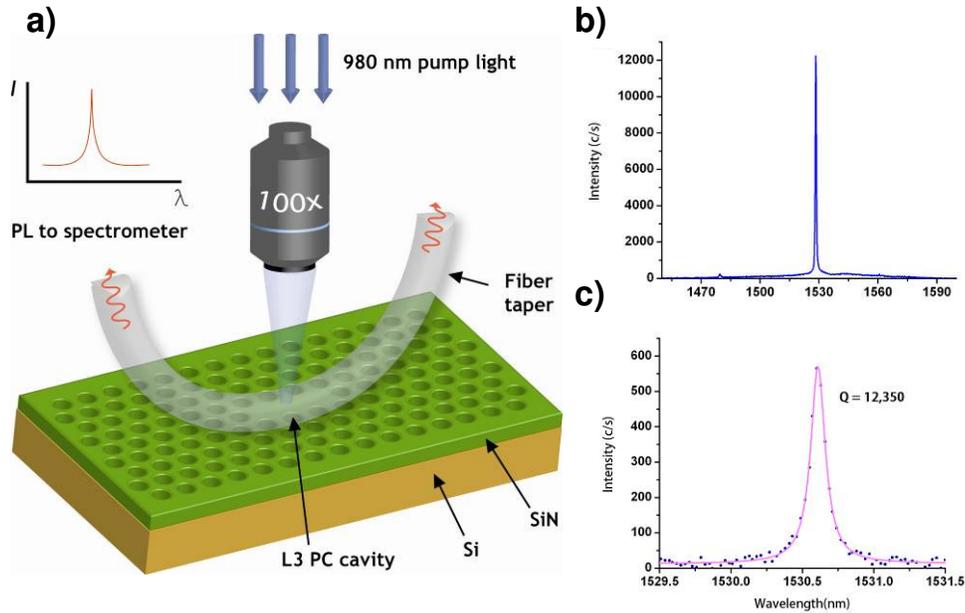

Fig. 6. (a) Experimental setup for outcoupling PL from Er:SiNx on Si PC cavities via fiber tapers. Pump light is focused on the cavity while PL emission is collected by the same fiber taper and sent to a spectrometer. (b) Full PL spectrum for fiber-coupled emission from the PC cavity. The integrated intensity is 2.5x larger than for the free space measurement and the peak has a Q of 4,300. (c) A different data point for which the taper was offset from the cavity so that the cavity was minimally loaded.

When the taper is positioned over the center of the cavity, the PL extraction is maximized as can be seen in Fig. 6(b). The non-resonant background PL of erbium is minimized in this configuration while the integrated intensity of the fundamental mode is 2.5 times that collected through free space (in the vertical direction) and an NA = 0.5 objective lens (Fig. 4). Accounting for the loss in this particular taper and adding up the intensity in both arms, the total collection into the fiber is 12.5 times the collection through the NA = 0.5 lens in the direction perpendicular to the PC slab. Since the radiation calculation revealed that collection through free space is 4.2%, the total fiber coupling efficiency is 53%. This value is pretty close to the result predicted by simulation and additional discrepancies could be due to differences in the actual taper size and exact taper positioning. Such a high extraction efficiency is made possible because the fiber taper relies on evanescent coupling from the cavity and is not limited geometrically by the finite extent of an objective lens. It is interesting to note that a large coupling ratio can be obtained despite the phase matching between the taper and cavity not being fully optimized [19]. This is a result of the fact that a strongly localized cavity resonance spans a number of k-vectors, which permits coupling to a taper mode. Additionally, the curved taper itself spans a wide k-space [20] as was found out from separate coupling experiments with waveguides.

Fig. 6(c) shows the PL of the cavity resonance for the case when the taper is displaced from the cavity main axis. The Q in this case is found to be approximately 12,350, which corresponds to 98% of the intrinsic Q value. Coupling to fiber tapers in direct contact measurements is usually associated with high loading of the cavities due to either direct coupling into the fiber or parasitic loss into other decay paths [19]. As expected, we confirm that by reducing the coupling strength (by decreasing the mode field overlap), the cavity Q can remain very high. This agrees with simulation results in Fig. 3. The ability to have a control over the cavity Q is important for PC cavity-based lasers or amplifiers which need to have as low loss as possible.

To more clearly see the relationship between collection efficiency and measured Q value, a series of spectra were taken for various taper offsets from the cavity as seen in Fig. 7(a). The fiber taper was slowly displaced along y-direction and as expected, the measured cavity Q increased while the collected intensity decreased. Fig. 7(b) shows a few traces at different coupling strengths where the intensity-Q relationship is seen clearly. This behavior can be described by an efficiency relationship under the condition that no parasitic loading of the cavity takes place [9]:

$$\eta = 1 - \frac{Q_t}{Q_0} \qquad (1)$$

Here $\eta$ is the collection efficiency given as the decay rate into the fiber divided by the total decay rate, $Q_t$ is the loaded cavity Q, and $Q_0$ is the intrinsic Q. The straight line in Fig. 7(a) is a fit to the data using Eq. (1). As can be seen in the plot, the fit is only good for the region of low coupling (high Q) and deviates significantly where taper loading is strong. This is expected from simulations, since for high coupling the parasitic Q grows and the denominator in equation (1) would have to be modified by adding a parasitic $Q_p$ term. The additional degradation is due to the newly generated decay from the parasitic TM coupling.

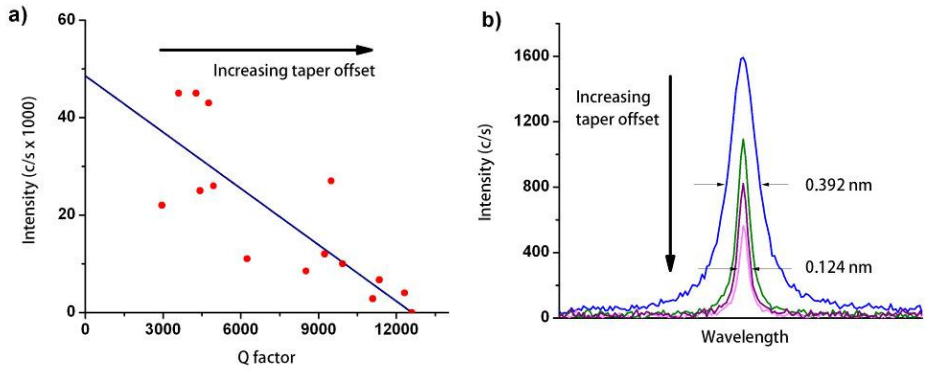

Fig.7. (a) Collected PL intensity and the total cavity Q-factor as the fiber taper is offset from the cavity along the y-direction. The blue line is a fit to the data. (b) Several spectra for the data in (a). As the taper moves away from the cavity the integrated intensity decreases and the Q increases. The peaks are centered together for easier visualization.

## 5. Conclusion

In summary, we demonstrate fiber taper coupling and efficient PL extraction from a CMOS-compatible light-emitting material, Er:SiNx embedded in a Si PC cavity. The collection efficiency through the fiber taper was 2.5 times the collection through free space and could be improved up to 6.25 times through better taper fabrication and even 12.5 times with an interferometric taper design [21]. Larger collection enhancements are also possible by better phase matching of the fiber and cavity modes. The loaded quality factor of the coupled cavity can remain very high and in the limit of weak coupling can be as high as 98% of its intrinsic value. A broad tuning range of collection efficiencies and quality factors is made possible by a variable coupling strength inversely proportional to the taper displacement from the cavity. The fiber taper provides a convenient platform for linking the nano-scale on-chip environment with the off-chip interconnect network, which is important for building devices such as on-chip optical amplifiers. For the promising erbium-doped silicon nitride material, fiber coupled emission is an easy and efficient way to extract light and connect to outside systems.

**Acknowledgements**

The authors would like to acknowledge the MARCO Interconnect Focus Center and the U.S. Air Force MURI program under Award No. FA9550-06-1-0470 for funding. Gary Shambat and Yiyang Gong would also like to thank the NSF GRF for support. The fabrication has been performed in the Stanford Nanofabrication Facilities.